\documentclass[a4paper]{jpconf}
\usepackage{graphicx}
\usepackage{amsmath}

\begin{document}
\title{Non-equilibrium fluctuations at the QCD phase transition}

\author{M. Nahrgang, M. Bleicher}

\address{Institut f\"ur Theoretische Physik \& Frankfurt Institute for Advanced Studies (FIAS), Goethe-Universit\"at, Max-von-Laue-Str.~1, 
D-60438 Frankfurt am Main, Germany}

\ead{nahrgang@th.physik.uni-frankfurt.de}

\begin{abstract}
We study the chiral phase transition by the non-equilibrium propagation of the sigma field. A quark fluid acts as a heat bath in local thermal equilibrium and evolves fluid dynamically. We allow for dissipative processes and fluctuations since the sigma field is propagated according to a Langevin-type equation of motion. 
Non-equilibrium fluctuations at the first order phase transition lead to an increase in the intensity of sigma excitations.
\end{abstract}

\section{Introduction}
The QCD phase transition is subject to intensive theoretical and experimental research. Besides theoretical indications of a QCD critical point \cite{Fodor:2001pe,Schmidt:2008ev}
experimental signatures are proposed to discover the suggested critical point in experiment \cite{Stephanov:1998dy,Stephanov:1999zu}. 
The fast dynamics during the expansion of the system created in a collision makes it important to consider non-equilibrium phenomena.
At a critical point relaxation times become infinite. A system that cools rapidly through a second order phase transition is necessarily driven out of equilibrium. Any signal for a critical point is crucially weakened \cite{Berdnikov:1999ph}. A non-equilibrium situation, however, leads to interesting phenomena at a first order phase transition, such as nucleation \cite{Chomaz:2003dz} and spinodal decomposition \cite{Mishustin:1998eq}. 
We embed a field theoretical model with a phase transition into a realistic expansion of the hot and dense matter and
extend similar models \cite{Mishustin:1998yc,Paech:2003fe} by a complete treatment of fluctuations and dissipation of the chiral order parameter coupled to the fluid dynamic expansion. While the chiral field is damped energy dissipates into the quark fluid. Random kicks by the noise term also contribute to the energy and momentum exchange between the field and fluid. We introduce a stochastic source term into the fluid dynamic expansion to conserve energy of the whole system. 
\section{Chiral fluid dynamics}
\subsection{Semiclassical equation of motion for the sigma field including dissipation and noise terms}
We use the linear sigma model with consituent quarks as a low energy effective model of QCD. It exhibits a chiral phase transition at different strengths, a crossover, a critical point and a first order phase transition \cite{Scavenius:2000qd}
\begin{equation}
{\cal L}=\overline{q}\left[\rm{i}\gamma\partial_\mu-g\left(\sigma+\text{i}\gamma_5\tau\vec{\pi}\right)\right]q 
  + \frac{1}{2}(\partial_\mu\sigma\partial^\mu\sigma)+
 \frac{1}{2}(\partial_\mu\vec{\pi}\partial^\mu\vec{\pi}) 
- U(\sigma, \vec{\pi}) \, ,
\label{eq:LGML}
\end{equation}
with the constituent quark field $q=\left(u,d\right)$, the coupling $g$ between the quarks and the chiral fields. The potential is given by
\begin{equation}
U\left(\sigma, \vec{\pi}\right)=\frac{\lambda^2}{4}\left(\sigma^2+\vec{\pi}^2-\nu^2\right)^2-h_q\sigma-U_0\, ,
\label{eq:Uchi}
\end{equation} 
where $\langle\sigma\rangle=f_\pi=93$~MeV and $\langle\vec\pi\rangle=0$. The explicit symmetry breaking term is $h_q\sigma=f_\pi m_\pi^2$ with $m_\pi=138$~MeV. Thus $\nu^2=f_\pi^2-m\pi^2/\lambda^2$. $\lambda^2=20$ yields a sigma mass $m_\sigma^2=2\lambda^2 f_\pi^2 + m_\pi^2\approx 604$ MeV. For zero potential energy in the ground state $U_0=m_\pi^4/(4\lambda^2)-f_\pi^2 m_\pi^2$.

In the rest of this paper we will concentrate on the order parameter of the chiral phase transition, the sigma field and set $\pi=\langle\pi\rangle=0$ for all times.
The equation of motion for the sigma field including dissipation and noise is given by
\begin{equation}
 \partial_\mu\partial^\mu\sigma+\frac{\delta U}{\delta\sigma}+g\langle\bar{q}q\rangle_\sigma +\eta\partial_t\sigma=\xi\, .
\label{eq:eomsigma}
\end{equation}
Here, the chiral condensate to one-loop level is
\begin{equation}
 \langle\bar{q}q\rangle_\sigma=2g d_q\sigma\int\frac{{\rm d}^3p}{(2\pi)^3}\frac{1}{E}n_{\rm F}(E)
\end{equation}
$n_{\rm F}(p)$ is the Fermi-Dirac distribution, $d_q=12$ the degeneracy factor and $E=\sqrt{p^2+g^2\sigma^2}$ the energy of the quarks. 

Our choice of the damping coefficient $\eta$ and the noise term $\xi$ is motivated by the results of two-loop calculations in a simplified ${\cal O}(4)$ chiral model \cite{Xu:1999aq,Greiner:1996dx}. In a first approximation we assume that $\eta$ is independent of temperature and mass changes and that the noise is white. Inspired by \cite{Biro:1997va} we use $\eta=2.2/{\rm fm}$. The dissipation-fluctuation theorem then gives
\begin{equation}
 \langle \xi(t)\rangle=0\quad {\rm and}\quad \langle\xi(t)\xi(t')\rangle=\frac{2T}{V}\eta\delta(t-t')\; .
\end{equation}

\subsection{The effective potential and the equation of state}
The pressure of the quark fluid is given by the thermodynamic potential. To one-loop level it is
\begin{equation} 
V_{{\rm eff}}(\sigma, \vec{\pi},T)=-\frac{T}{V}\log {\cal Z} =-2d_q T \int\frac{{\rm d}^3p}{(2\pi)^3}\log(1+{\rm e}^{-\frac{E}{T}}) + U\left(\sigma, \vec{\pi}\right)\; .
\end{equation}
At $\mu_B=0$ the strength of the phase transition can be tuned by varying the strength of the coupling $g$.  For a small coupling the transition is a crossover, for $g=3.63$ the potential becomes flat as for a continuous transition. And for $g=5.5$ the phase transition is discontinuous. 

The equation of state for the fluid dynamic expansion is obtained from 
\begin{eqnarray}
  p(\sigma, \vec{\pi},T)&=& -V_{\rm eff}(\sigma, \vec{\pi},T)+U(\sigma, \vec{\pi})\\
  e(\sigma, \vec{\pi},T)&=& T\frac{\partial p(\sigma, \vec{\pi},T)}{\partial T}-p(\sigma, \vec{\pi},T)\; .
\end{eqnarray}

\subsection{The stochastic source term}
The equations of relativistic fluid dynamics are
\begin{equation}
\partial_\mu (T_{\text{fluid}}^{\mu\nu}+T_{\text{field}}^{\mu\nu})=0
\label{eq:fluidT}
\end{equation}
By using the explicit form of the equation of motion (\ref{eq:eomsigma}) we derive for the source term
\begin{equation}
S^\nu=-\partial_\mu T_{\text{field}}^{\mu\nu}=-(\partial_\mu\partial^\mu\sigma+\frac{\delta U}{\delta\sigma})\partial^\nu\sigma
=-(-g\langle\bar q q\rangle_\sigma-\eta\partial_t\sigma+\xi)\partial^\nu\sigma
\label{eq:sourceterm}
\end{equation}
The damping and the noise term enter the source term which is therefore also stochastic. 

\section{Numerical results}
We solve the fluid dynamic equations (\ref{eq:fluidT}) by a SHASTA code and the field equations (\ref{eq:eomsigma}) by a staggered leap-frog algorithm. The numerical equilibrium solution of equations like (\ref{eq:eomsigma}) depends on the lattice spacing \cite{CassolSeewald:2007ru}. It turns out, however, that the results presented here depend only slightly on the lattice spacing, for which we use $\Delta=0.2$ fm.
The energy density is initiated as the equilibrium energy density at $T=160{\rm MeV}$, ellipsoidal in x-y-plane and homogenous in z-direction, smoothed by a Woods Saxon type distribution.
 The sigma field is initiated in equilibrium with the quark fluid. All observations shown below are purely due to non-equilibrium effects at the phase transition.

\begin{figure}[h]
\begin{minipage}{18pc}
\includegraphics[width=13pc,angle=-90]{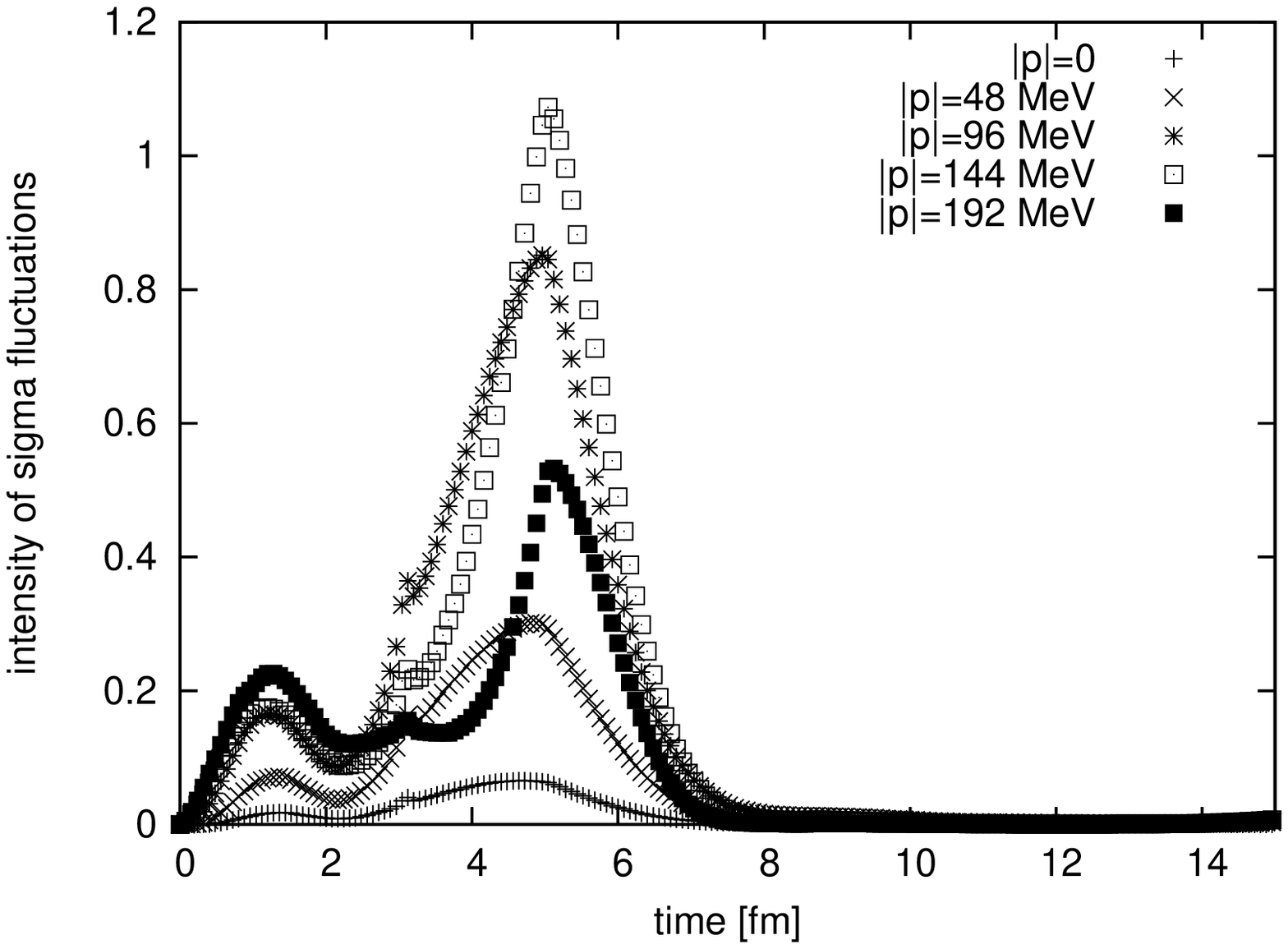}
\caption{\label{fig:signumcp}Time evolution of the intensity of sigma fluctuations at a critical point.}
\end{minipage}\hspace{1pc}%
\begin{minipage}{18pc}
\includegraphics[width=13pc,angle=-90]{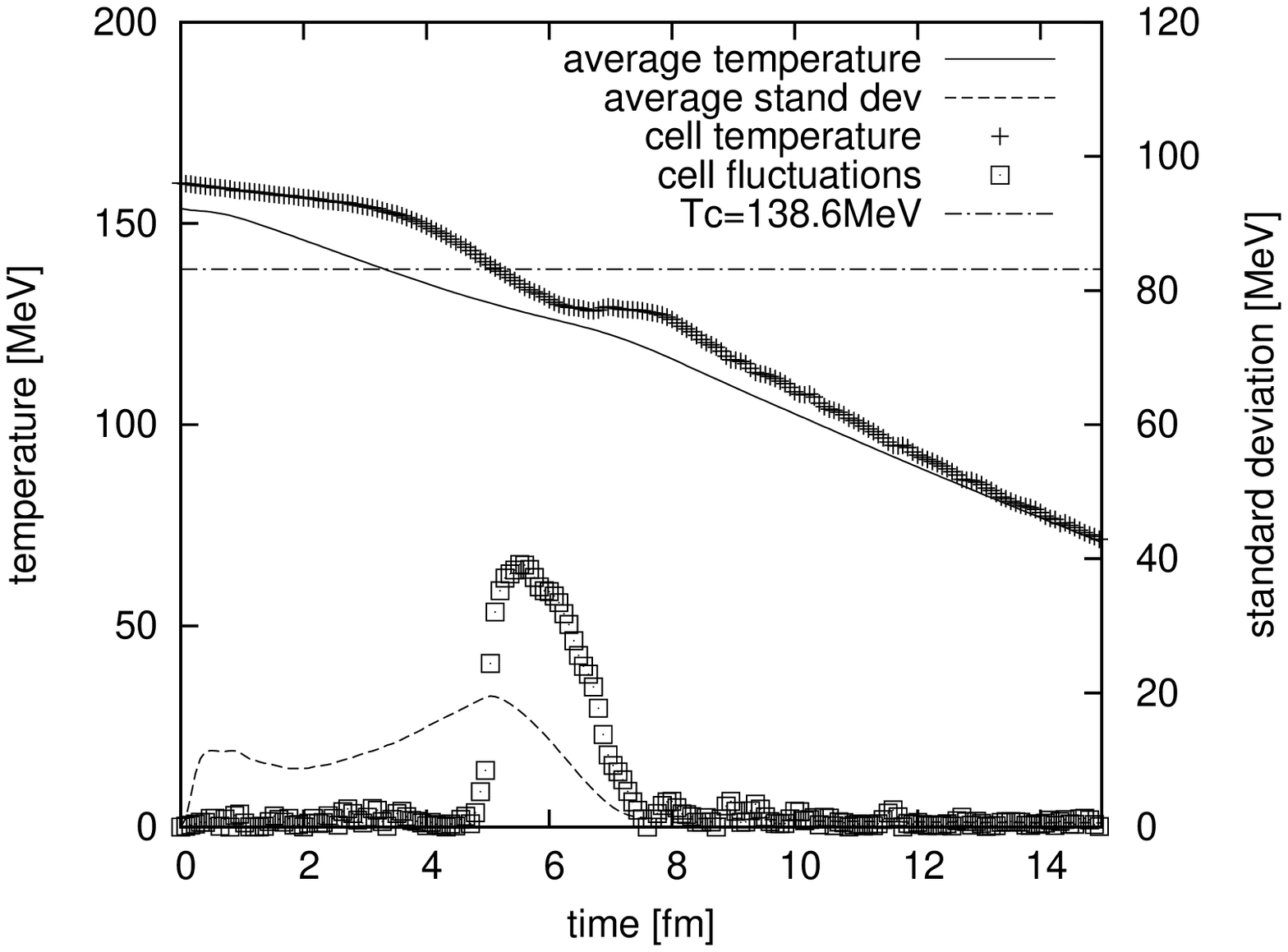}
\caption{\label{fig:fluccp}Temperature of the quark fluid and standard deviation of the sigma field at a critical point.}
\end{minipage} 
\end{figure}

The intensity of the sigma fluctuations is given by
\begin{equation}
 \frac{{\rm d}N_\sigma}{{\rm d}^3k}=\frac{a_k^\dagger a_k}{(2\pi)^3 2\omega_k}=\frac{1}{(2\pi)^3 2\omega_k}{(\omega_k^2|\sigma_k|^2+|\dot\sigma_k|^2)}
\label{eq:numsig}
\end{equation}
\begin{figure}[h]
\begin{minipage}{18pc}
\includegraphics[width=13pc,angle=-90]{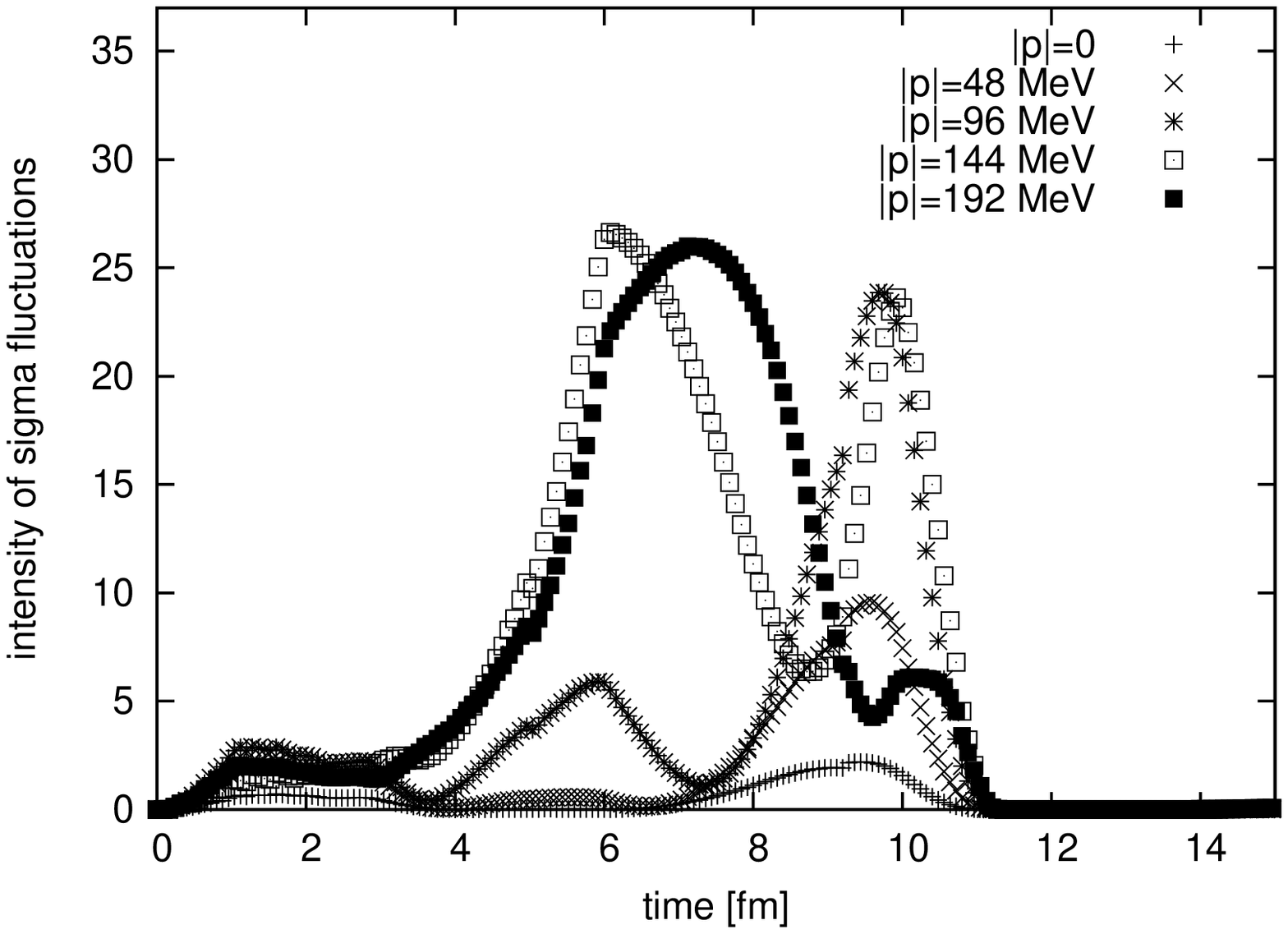}
\caption{\label{fig:signumfo}Time evolution of the intensity of sigma fluctuations at a first order phase transition.}
\end{minipage}\hspace{1pc}%
\begin{minipage}{18pc}
\includegraphics[width=13pc,angle=-90]{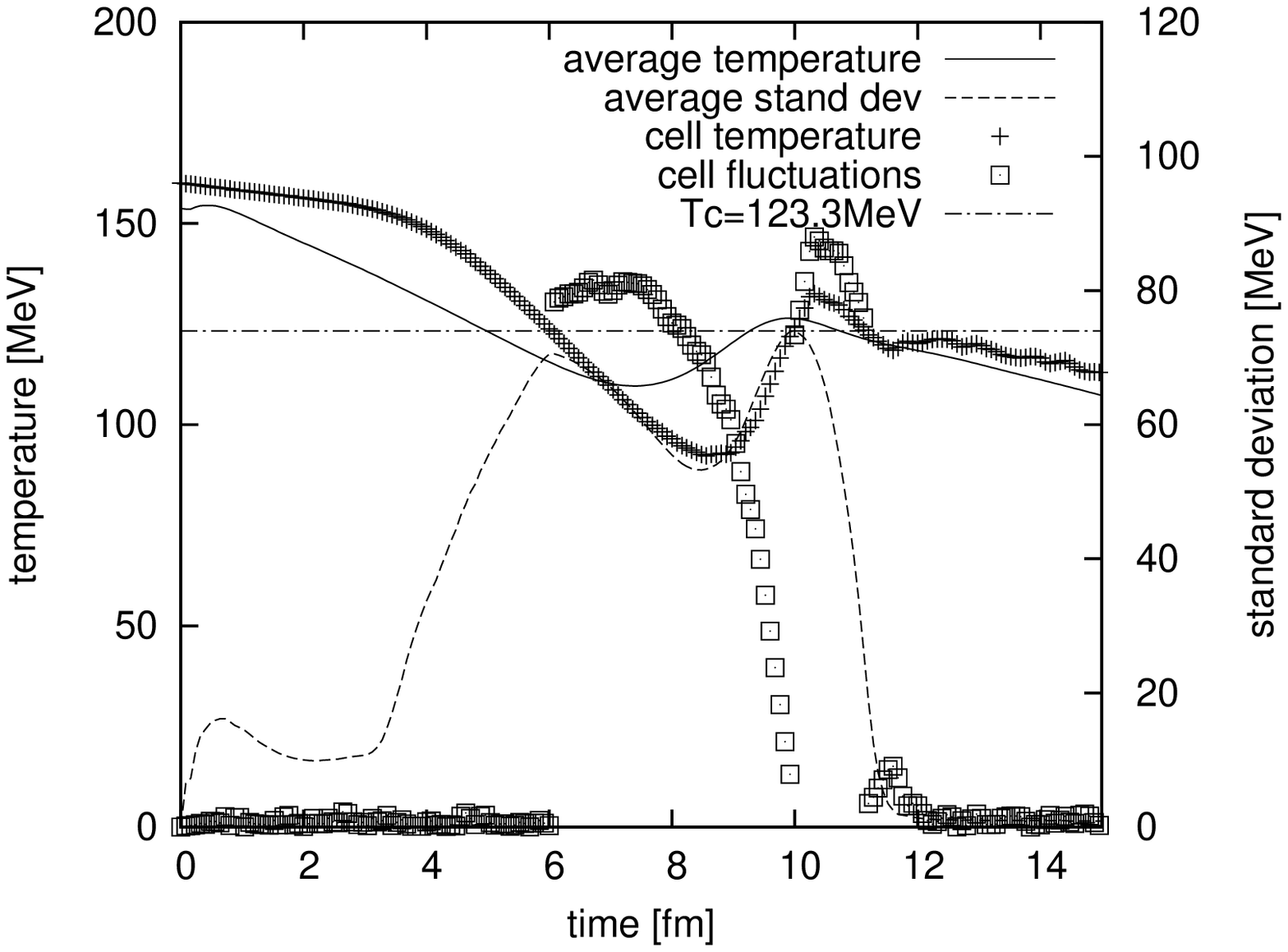}
\caption{\label{fig:flucfo}Temperature of the quark fluid and standard deviation of the sigma field at a first order phase transition.}
\end{minipage} 
\end{figure}
The time evolution of (\ref{eq:numsig}) is shown in figure (\ref{fig:signumfo}) for a first order phase transition and in figure (\ref{fig:signumcp}) for a critical point. In a first order phase transition the intensity of sigma fluctuations between $4$ and $11$ fm is much larger than at a critical point. 
This can be better understood by looking at the evolution of the temperature and the size of the fluctuations. We show these values in a single cell in the middle of the grid and averaged over a volume in the hot region of the fireball ($V=0.24$ fm$^3$). The standard deviation of the fluctuations of the sigma field in V is $\Delta\sigma^2=1/V\sum_V\Delta ^3(\sigma_{\rm eq}-\sigma(x))^2$. For a first order phase transition the intensity of sigma fluctuations increases around $t=5$ fm when the average temperature drops below the critical temperature, in figure (\ref{fig:flucfo}). Here the standard deviation increases to almost $70$ MeV. In one particular cell in the middle of the hot region one sees that the size of the fluctuation corresponds to the difference between the two minima at $\langle\sigma\rangle_{T>T_C}\simeq10$ MeV and $\langle\sigma\rangle_{\rm vac}=f_\pi=93$ MeV.
Obviously the fluctuations in a critical point scenario, figure (\ref{fig:fluccp}), are smaller.
For a first order phase transition we find a second rise in the intensity of sigma fluctuations at around $t=10$ fm. While the sigma field relaxes from the unstable minimum towards its vacuum value energy is transfered to the quark fluid where re-heating occurs. This effect is so strong that the average temperature again crosses the critical temperature leading to an increase in the fluctuations. Thereafter the sigma field relaxes slowly towards $\langle\sigma\rangle_{\rm vac}$.

\section{Summary}
We presented a dynamic model of phase transitions in heavy ion collisions. We included damping and noise originating from the interaction of the chiral fields with the quark fluid. To ensure energy and momentum conservation of the coupled system we suggested to introduce a stochastic source term into the equations of fluid dynamic. 
The number of coherently produced soft sigmas is much larger at a first order phase transition than compared to a critical point transition. Non-equilibrium effects like supercooling and reheating are observed.

\subsection{Acknowledgments}
 The authors thank Carsten Greiner, Stephan Leupold, Igor Mishustin and Christoph Herold for fruitful discussions and Dirk Rischke for providing the hydro code.
 This work was supported by the Hessian LOEWE initiative Helmholtz International Center for FAIR. 
 M.~Nahrgang acknowledges finanical support from the Stiftung Polytechnische Gesellschaft Frankfurt.

\end{document}